\documentstyle[prc,epsf,multicol,aps]{revtex}

\begin{document} 
\draft 

\title{Competition of isoscalar and isovector proton-neutron pairing
  in nuclei}

\author{G. Mart\'{\i}nez-Pinedo$^1$,  K. Langanke$^1$ and
  P. Vogel$^2$}   

\address{$^1$Institute for Physics and Astronomy and 
  Theoretical Astrophysics Center\\
  University of Aarhus, DK-8000 Aarhus C, Denmark\\
  $^2$Physics Department, California Institute of Technology,
  Pasadena, CA 91125, USA}

\date{\today} 

\maketitle

\begin{abstract} 
  We use shell model techniques in the complete $pf$ shell to study
  pair correlations in nuclei. Particular attention is paid to the
  competition of isoscalar and isovector proton-neutron pairing modes
  which is investigated in the odd-odd $N=Z$ nucleus $^{46}$V and in
  the chain of even Fe-isotopes.  We confirm the dominance of
  isovector pairing in the ground states. An inspection of the level
  density and pair correlation strength in $^{46}$V, however, shows
  the increasing relative importance of isoscalar correlations with
  increasing excitation energy.  In the Fe-isotopes we find the
  expected strong dependence of the isovector pairing strength on the
  neutron excess, while the dominant $J=1$ isoscalar pair correlations
  scale much more gently with neutron number. We demonstrate that the
  isoscalar pair correlations depend strongly on the spin-orbit
  splitting.
\end{abstract} 


\begin{multicols}{2}

\section{Introduction and motivation}

Stimulated by progress in many-body theories and the advent of novel
experimental tools like the radioactive ion-beam facilities, pairing in
nuclei, and in particular proton-neutron ($pn$) pairing, has become
quite fashionable recently~\cite{Engel96,Langanke97,Satula97,Civitarese97,Dobes97,Poves98,Engel97,Engel98,VanIsacker97,Vogel98}.
In contrast to like-nucleon pairing, $pn$ pairing may exist in two
distinct varieties - isovector and isoscalar pairing.  Isovector $pn$
pairing has been found to decrease fast with increasing neutron excess
(or more precisely with increasing $|N-Z|$, where $N,Z$ are the
numbers of neutrons and protons, respectively) within an isotope
chain~\cite{Engel96}. Thus, it is obvious that $pn$ pairing is
relatively more important in $N=Z$ nuclei. While even-even $N=Z$
nuclei have ground state isospin $T=0$ and isospin symmetry forces
isovector pairing to be identical in all three $T=1$ pairing channels
(proton-proton ($pp$), neutron-neutron ($nn$), and proton-neutron),
this symmetry is broken in most odd-odd $N=Z$ nuclei with mass number
$A>40$ which have $T=1$ ground states.  Indeed, it is observed that
isovector $pn$ pairing is the dominating pairing mode in these nuclei.
On the other hand, odd-odd $N=Z$ nuclei with $A<40$ have ground state
isospin $T=0$ (with the only exception of $^{34}$Cl) pointing to a
competition between the two different $pn$ pairing modes and its
change with mass number is an obvious source of interest. The
interplay of isovector and isoscalar $pn$ pairing has been studied in
various contexts:

\begin{itemize}
\item It has long been known that $N=Z$ nuclei gain extra binding
  compared to the smooth trend defined by their neighbors.  This extra
  piece, called the Wigner energy, is stronger for even-even than for
  odd-odd nuclei. Recent studies of  the Wigner
  energy tried to relate its origin to the $pn$ pairing \cite{Satula98}.

\item Odd-odd $N=Z$ nuclei are the only ones with two nearly
  degenerate low-lying isospin states ($T=0$ and $T=1$). This fact and
  the change of the ground state isospin from $T=0$ to $T=1$ at $A=40$
  has been explained by a near balance of isovector pairing
  and symmetry energy \cite{Vogel98}.

\item While the ground state isospin of odd-odd $N=Z$ nuclei with
  $A>40$ (except for $^{58}$Cu) is $T=1$, it is experimentally found
  that the density of $T=0$ levels at modest excitation energies is
  larger than the density of $T=1$ states. For example, for $^{74}$Rb a
  transition from isovector to isoscalar dominance has recently been
  observed with increasing rotational frequency \cite{Rudolph}. This
  transition has been traced back to a competition between isovector
  $J=0$ $pn$ pairing and isoscalar aligned $pn$ pairing ($J=9$), where
  the latter becomes obviously more important at larger angular
  momenta~\cite{Dean97}.

\end{itemize}

The interplay and competition of isovector and isoscalar pairing is
also the theme of this paper, where we study and discuss two different
aspects of this topic. At first, we perform detailed studies of the
spectrum of $^{46}$V. Special emphasis is paid to the comparison of
the $T=0$ and $T=1$ parts of the spectra and their dependence on the
pair correlation strength. In particular, we investigate the
temperature dependence of the various pairing modes tracing possible
differences in the behavior of the isovector and isoscalar $pn$
pairing.

Isovector pairing in the dominating $J=0$ channel involves coupling of
states in time-reversed components of the same orbitals. Very often
isovector pairing is effectively given by one single matrix element
describing the $J=0$ pairing of the valence nucleons in one partly
filled orbital (the $f_{7/2}$ orbital in the nuclei studied in this
paper). In contrast, isoscalar $pn$ pairing, even its $L = 0$
component, can involve not only pairing within the same orbital, but
also between spin-orbit partners.  Thus isoscalar $pn$ pairing
involves several scales: the coupling matrix element(s) in the same
orbital, the usually strong (e.g. $J=1$) matrix element between spin-orbit
partners, and finally the spin-orbit splitting which can be understood
as a penalty which reduces this otherwise very favorable correlation.
To shed some light on the importance of isovector versus isoscalar
pairing, and in particular, on the role played by the spin-orbit
splitting we study pair correlations within the even Fe-isotopes
$^{50-56}$Fe comparing the results of ``realistic calculations'' to
those in which we have artificially reduced the spin-orbit splitting.

Our studies have become feasible due to recent progress made in
solving the interacting shell model, which is the method of choice to
investigate correlations among nucleons. All calculations are
performed with the realistic KB3 interaction~\cite{Poves81} in the
complete $pf$ shell.  Modern diagonalization codes allow now studies
in model spaces whose dimensions have been considered untractable only
a few years ago. We use such a state-of-the-art diagonalization
code~\cite{Antoine} in our calculation of $^{46}$V and hence have the
possibility to study the spectrum and its properties in a
state-by-state approach. About a year ago, when we started our
investigations of pairing properties in the Fe-isotopes,
diagonalization codes seemed not to be able to handle valence model
spaces of these dimensions (for very recent progress, the reader is
refered to~\cite{Caurier98}).  For that reason we have adopted the
Shell Model Monte Carlo (SMMC) approach \cite{Johnson,Lang,physrep}
for these studies. In the SMMC the nucleus is described by a canonical
ensemble at finite temperature. Since the Monte Carlo techniques avoid
an explicit enumeration of the many-body states, they can be used in
model spaces far larger than those accessible to conventional
diagonalization methods. As a disadvantage compared to these methods,
the SMMC approach cannot study detailed spectroscopy and ground state
properties are achieved as the limit of low temperatures. To
circumvent the ``sign problem'' encountered in the SMMC calculations,
we adopt the ``g-extrapolation'' procedure suggested in Ref.
\cite{sign}.

Both methods - the diagonalization approach and the SMMC - are well
documented in the literature and thus there is no need for a
repetition here. For detailed descriptions of the methods we refer the
reader to~\cite{physrep,Caurier94}.

\section{Results and discussion}

\begin{figure}
  \begin{center}
    \leavevmode
    \epsfxsize=0.8\columnwidth
    \epsffile{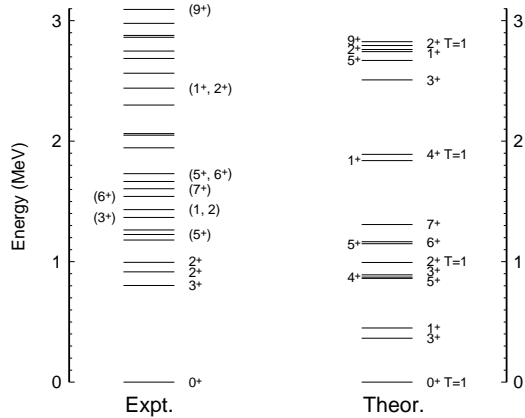}
    \caption{Calculated and experimental spectra of $^{46}$V. Only the
      isospin of the ground state is known experimentally. The
      energies of the excited T=1 states in $^{46}$V can be estimated
      from the isobaric analog states in $^{46}$Ti: 2$^+$ (0.889~MeV),
      4$^+$ (2.010~MeV), 0$^+$ (2.611~MeV), 2$^+$ (2.962~MeV).}
    \label{fig1}
  \end{center}
\end{figure}

First we discuss results obtained for the odd-odd $N=Z$ nucleus
$^{46}$V by diagonalization in the complete $pf$ shell. The calculated
spectrum is compared to the experimental spectrum~\cite{nds} in
Fig.~\ref{fig1}. We find overall good agreement, although the
calculated $T=0$ spectrum is at slightly too low (by about 400~keV)
excitation energy.

Given the spectrum of the nucleus, we define the internal energy at a
finite temperature $T$ by ($\beta=1/T$)

\begin{equation}
U(T)= \frac{1}{Z}
\sum_i (2 J_i+1) E_i \exp[{-\beta E_i}] 
\end{equation}
with the partition function
\begin{equation}
Z=\sum_i (2J_i+1) \exp[{- \beta E_i}] \; ;
\end{equation}
the sums runs over all nuclear levels $i$ labeled by quantum numbers
($T_i,J_i$) and excitation energy $E_i$. Fig. 2 displays the internal
energy and the specific heat $C_v= \frac{d}{dT} U(T)$ as function of
temperature $T$. The striking feature is the double-peak structure in
the specific heat. While the peak around $T=1.5$ MeV is trivially
related to the finite size of our model space (Schottky effect), the
peak at $T=0.2$ MeV is a unique structure of odd-odd $N=Z$ nuclei with
$A>40$ (except for $^{58}$Cu). As we pointed out earlier, these nuclei
have ground state isospins $T=1$ and $J=0$, and exhibit, like their
even-even analogs $^{46}$Ti and $^{46}$Cr, a sparse spectrum of $T=1$
levels at low energies reflecting the strong isovector pairing.
However, in odd-odd $N=Z$ nuclei the lowest levels with $T=0$ and
$T=1$ are almost degenerate reflecting a competition of pairing and
symmetry energy.  Furthermore, as can be seen in Fig. 1, odd-odd $N=Z$
nuclei exhibit several $T=0$ levels at low excitation energy and these
levels have $J>0$.

The peak in the specific heat at $T=0.2$~MeV signals the
``transition'' from the $T=1$ ground state to the bunch of $T=0$
levels (with larger statistical weight) at low excitation energy. We
note that the peak will be somewhat shifted towards higher
temperatures if we add an attractive component $b{\hat T}^2$, with the
isospin operator ${\hat T}$, to the Hamiltonian to correct for the
slight mismatch between the $T=1$ and $T=0$ levels in our calculation.
In passing we also remark that such a peak is not visible in the
specific heat for odd-odd $N=Z$ nuclei with $T=0$ ground states and
angular momenta $J>0$, where the $T,J=1,0$ level is at a modest
excitation energy and due to its small statistical weight does not
show up as peak in the specific heat.

To complete our discussion of the $^{46}$V spectrum we have plotted
the level density in Fig. 3. For clarity we have binned the results in
0.5 MeV bins and exhibit them for both isospin channels $T=0$ and
$T=1$ separately. As is customary, we compare our shell model level
densities with the backshifted Fermi gas~\cite{Thielemann}.
\begin{equation}
\rho (U) = \frac{\sqrt{\pi}}{12 a^{1/4}}
\frac{\exp{2 \sqrt{aU}}}{U^{5/4}}  \; ; \; U=E-\Delta
\end{equation}
where $\Delta,a$ are the backshift and level density parameter,
respectively.  These two essential parameters of the model are
determined by global fits to experimental data at low energies (around
the neutron threshold and below), leading to the approximate
expressions~\cite{Thielemann}
\begin{eqnarray}
a & \approx & A/8 \; {\rm MeV}^{-1}  \\
\Delta & \approx & \pm \frac{12}{\sqrt{A}} - \frac{10}{A} \; {\rm MeV} 
\end{eqnarray}
with the positive (negative) sign for even-even (odd-odd) nuclei and
$\Delta=0$ for odd-$A$ nuclei. We note that the experimentally
determined value of $a$ is significantly larger than the Fermi gas
value ($a \approx A/16$ MeV$^{-1}$) indicating the presence of
additional correlations in the low-energy levels other than those
described by pairing.  The calculated $^{46}$V level density is
reasonably well described in the energy regime $E>2.5$ MeV by the
backshifted Fermi gas model with the parameters $\Delta=-2.0$ MeV and
$a=3.5$ MeV$^{-1}$.  While $\Delta$ agrees well with the empirical
value, our $a$ is significantly smaller suggesting the presence of
states outside of our model space in $^{46}$V already at rather low
excitation energies. The discrepancy in $a$ is less than the
comparison with Eq. (4) suggests; in Ref.~\cite{Woosley} the values
$\Delta = -1.79$~MeV and $a$ = 4.9 MeV$^{-1}$ are recommended, closer
to the ones found here.

As we have seen already in the discussion of the spectrum, pairing
plays an essential role. We will now investigate this topic in more
detail by studying the pair correlation strength in $^{46}$V as a
function of excitation energy. We restrict our discussion to
$s$-wave pairing only, hence $L=0$. As a measure of the pair
correlation strength we use the operator ${\cal N}_{JTt_z}$, defined
in $LS$-coupling by:
\begin{equation}
{\cal N}_{JTt_z} = 
\sum_{ll'M} \sqrt{(2l+1)(2l'+1)} A_{0SJM}^{Tt_z \dagger}(ll)
A_{0SJM}^{Tt_z}(l'l') 
\label{e:pair}
\end{equation}
with the two particle creation operator $A^{\dagger}$:
\begin{equation}
A_{LSJM}^{ Tt_z \dagger }(l_A l_B) =
\frac{1}{\sqrt{1 + \delta_{AB}}}
\left[a_A^\dagger a_B^\dagger\right]_{LSJM}^{Tt_z} .
\end{equation}
and $a_{nlms_zt_z}^\dagger$ creates a nucleon of isospin projection
$t_z$ and spin projection $s_z$ in the orbital $nlm$. In our studies
involving the diagonalization approach the expectation value is
evaluated state-by-state. In the SMMC approach the expectation value
refers to a thermal average defined as
\begin{equation}
\langle {\cal O} \rangle = \frac{{\rm Tr_A} \exp[{- \beta H}] {\cal O}}{Z}
\end{equation}
where ${\rm Tr_A}$ is the canonical trace (fixed numbers of neutrons and
protons) and $H$ is the many-body Hamiltonian. 

In the limit of large degeneracy $\langle \cal N \rangle$ represents
the number of nucleon pairs with the angular momentum $J$, isospin $T$
and its projection $t_z$.  In $pf$ shell nuclei the degeneracies are
not too large, the operators $A, A^{\dagger}$ are not really bosons,
and hence $\langle {\cal N}_{t_z} \rangle$ can deviate from the true
pair number expectation value.

Fig.~\ref{fig4} shows the isovector $(T=1,S=0$) and isoscalar
$(T=0,S=1)$ pair correlation strength as function of excitation energy
(averaged in 0.5 MeV wide bins) and for $T=0$ and $T=1$ states in
$^{46}$V separately.  Isospin symmetry requires $\langle {\cal
  N}_{J=0,T=1,t_z} \rangle$ to be identical for all three isovector
channels ($pp,~nn,~pn$) in $T=0$ states, while isovector $pn$ pairing
can be different from like-particle pairing in the $T=1$ states ($pp$
and $nn$ pairing strengths are still identical).  Several observations
with respect to Fig.~\ref{fig4} are noteworthy: Isovector pairing
dominates in both $T=0$ and $T=1$ states at low excitation energies.
With the exception of the states below 1~MeV, isovector pairing
decreases, approximately exponentially, in $T=0$ states with energy.
On the other hand, the isoscalar pairing in $T=0$ states is about
constant up to $E=6$ MeV and then decreases more slowly than isovector
pairing. We conclude that isovector pairing in $T=0$ states is more
concentrated in the low-energy part of the spectrum, while isoscalar
pairing is also present at higher energies. The difference is, as
stated earlier and shown in detail below, presumably due to spin-orbit
splitting, which introduces an additional energy scale into the
isoscalar $pn$ pairing mode ($\epsilon_{7/2}-\epsilon_{5/2} \approx 6$
MeV) which is only overcome at higher energies.

The lowest $T=1$ states are dominated by isovector $pn$ pairing. In
these states, like-particle pairing is roughly constant, in fact it
increases slightly with energy.  We conclude that the isovector
pairing gap in $^{46}$V is due to $pn$ pairing. For states with $E^* >
4$ MeV, when the ``pairing gap'' is overcome, we find no large
difference between the three isovector pairing correlation strengths.
While the sum of isovector pairing is still larger than the isoscalar
pairing strength, the latter is stronger than each of the three
individual channels. As for $T=0$ states, isoscalar pairing decreases
more slowly than isovector pairing and becomes the largest correlation
for $E^*>15$ MeV (however, at these high excitation energies our model
space is quantitatively not adequate anymore and should be extended to
include the two neighboring major shells).

What is the origin of the pairing correlations? It is wellknown that a
``realistic'' shell model interaction can be approximated by a
monopole term and a two-body piece consisting of pairing and
quadrupole-quadrupole ($QQ$) terms (Bes-Sorensen
interaction~\cite{Bes}).  Thus the correlation can reflect both
genuine pairing interaction or deformation which is related to the
$QQ$ interaction.  To distinguish between these different sources we
define first the total correlation energy $H_{corr}$ by subtracting
the monopole contribution that includes both single particle energies
and average two body matrix elements~\cite{french} from the
expectation value of the full Hamiltonian. Next, we measure the
importance of the two dominating pieces in the residual interaction by
calculating the expectation values of the pairing, $H_{\text{pair}}$,
and quadrupole-quadrupole, $H_{QQ}$, hamiltonians in the eigenstates
of the full Hamiltonian. These Hamiltonians are defined by

\begin{mathletters}
\begin{eqnarray}
 H_{\text{pair}} = &&-G_{p01} \sum_{t_z} {\cal N}_{01t_z}\\
 \nonumber & & -G_{p10} {\cal N}_{100},
\end{eqnarray}
\begin{equation}
 H_{QQ} = - \chi \sum_\mu (-1)^\mu Q_\mu Q_{-\mu},
\end{equation}
where $Q_\mu$ is the mass quadrupole operator defined as:
\begin{equation}
 Q_\mu = \sqrt{\frac{16\pi}{5}} \sum_i^A r_i^2 Y_{2\mu}(\Omega_i).
\end{equation}
\end{mathletters}

The values of the different constants are taken from table II of
reference~\cite{zuker}.  After scaling to the corresponding
$\hbar\omega$ the values are: $G_{p01} = 0.37$~MeV,
$G_{p10}=0.57$~MeV, and $\chi = 0.014$~MeV. The expectation values are
shown in Fig. 5 as a function of temperature. We have additionally
split them into their isovector and isoscalar components defined by
taking the isovector and isoscalar pieces of the respective
Hamiltonians when they are written in the normal order.  The results
support the discussion given above.  The isovector correlation energy
is stronger than the isoscalar one. At the same time, while the
isoscalar correlation energy decreases gradually and slowly, there is
a noticeable drop in the isovector correlation energy at low
temperatures which can be traced back to the pairing. In the ground
state (temperature $T=0$) most of the correlation energy is due to
pairing energy, but the expectation value of the isovector component
of the pairing Hamiltonian has dropped to about half of its ground
state value already at temperature $T=0.25$ MeV.  By splitting
$H_{pair}$ into its various components, we verify that most of the
isovector pairing energy (about 60$\%$) and its drop with temperature
is caused by proton-neutron pairing.

Strikingly the isoscalar pairing correlation energy decreases very
slowly. While the results for the isovector pairing correlation energy
confirms the expectation and shows that this degree of freedom is
mainly concentrated in states at low excitation energies, our
calculation suggests that a well pronounced ``pairing gap'' does not
exist in isoscalar pairing. This conclusion is based on the comparison
of the isovector and isoscalar pair correlation energies in Fig. 5;
there is no peak at low energies in the $T=0$ channel. (The apparent
minimum of the $T = 0$ level density around 2 MeV in Fig. 3 appears to
be a statistical fluctuation.)

On the other hand, the isoscalar component of
$H_{QQ}$ increases slightly at small temperatures, counterbalancing
the slight decrease in the isoscalar piece of $H_{pair}$. The
isovector component of $H_{QQ}$ is roughly constant in the temperature
regime $T<0.25$ MeV, which is dominated by the states with large
isovector pairing contributions. We also notice that the isoscalar
quadrupole-quadrupole correlations decrease more slowly than the
isovector ones.

\begin{figure}
  \begin{center}
    \leavevmode
    \epsfxsize=0.8\columnwidth
    \epsffile{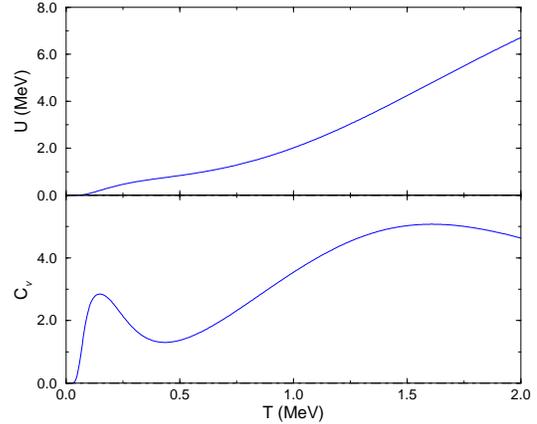}
    \caption{Internal energy $U$ (upper panel) and specific heat $C_V$
      (lower panel) of $^{46}$V as a function of temperature $T$.}
    \label{fig2}
  \end{center}
\end{figure}

In Figure~\ref{fig6} we compare the values of the different
correlations obtained by adding the isoscalar and isovector parts.
Clearly pairing plus quadrupole accounts for most of the total
correlation energy as measured by $H_{corr}$. The contributions of the
other components of the nuclear force (see~\cite{zuker} for a complete
characterization) are practically independent of the temperature. We
note that at $T\approx 0.15$~MeV, where the peak in the specific heat
in figure~\ref{fig2} appears, the quadrupole correlations became
dominant over pairing correlations as the low lying $T=0$ states, that
dominate the thermal average at this temperatures, have bigger
quadrupole energies than pairing energies.

Finally, we like to report on SMMC calculations of the even Fe-nuclei
$^{50-58}$Fe. We have chosen this set of isotopes as it includes an
$N=Z$ nucleus ($^{52}$Fe), a closed-shell nucleus ($^{54}$Fe), and
nuclei with valence protons and neutrons in different subshells
($^{56,58}$Fe). Our calculation parallels the SMMC study in
Ref.~\cite{Langanke97} where the reader might find details about the
approach.  Parts of the results concerning pairing have already been
presented in Ref.~\cite{Langanke97}, although in different context.

To distinguish between genuine pair correlations and those reflecting
the different number of neutrons, we introduce the excess of pair
correlations ${\cal N}_{exc}$ defined by subtracting the mean-field
values from the pair correlation strength defined in
Eq.~(\ref{e:pair}).  Following Ref.~\cite{physrep} we define the
mean-field value by the uncorrelated Fermi gas value using the
factorization
\begin{equation}
\langle a^\dagger_\alpha a^\dagger_\beta a_\gamma a_\delta \rangle =
n_\beta n_\alpha (\delta_{\beta\gamma}\delta_{\alpha\delta} -
\delta_{\beta\delta}\delta_{\alpha\gamma})
\end{equation}
where $n_\alpha$ are the occupation numbers.  Fig.~\ref{fig7} shows
the excess of isovector and isoscalar pair correlations.  Furthermore
we have indicated which orbital coupling mainly contribute to these
excesses.

Isovector $pn$ correlations are strongest in the $N=Z$ nucleus
$^{52}$Fe and decrease rapidly with increasing $|N-Z|$. This result
confirms the general trend already outlined in \cite{Engel96} which
noticed that with increasing $|N-Z|$ nuclear ground states show the
tendency to split into separate proton and neutron condensates. The
decreasing $pn$ correlations also allow an increase in $pp$
correlation with growing $|N-Z|$, as can be also seen in Fig. 7. The
excess of $nn$ correlations, however, does not follow the simple SO(5)
picture as shell effects, which are not present in that model, play an
essential role. One clearly observes that ${\cal N}_{exc}$ decreases
strongly towards the magic neutron number $N=28$ reflecting the
closure of the $f_{7/2}$ shell in $^{54}$Fe. With two neutrons outside
of $f_{7/2}$ neutron correlations increase again $(^{56}$Fe), but are
somewhat reduced again in $^{58}$Fe due to the partly closure of the
$p_{3/2}$ subshell. For this nucleus most of the $nn$ correlations are
due to pairing in the $f_{5/2}$ and $p_{1/2}$ orbitals.

A closer inspection of the $pp$ correlations indicates that the
$f_{5/2}$ orbitals contribute more than the $p_{3/2}$ orbitals,
although the latter are energetically favored. This is already a
signal for the presence of the isoscalar $pn$ pairing which introduces
$f_{7/2}-f_{5/2}$ correlations which are roughly the same for all
investigated Fe-isotopes. As long as the $f_{7/2}$ neutron shell is
not filled, we also find isoscalar $pn$ correlations among the
$f_{7/2}$ orbitals. For the heavier isotopes these correlations are
actually smaller than the mean-field values indicating that other
correlations become more important.

As suggested above, isoscalar $pn$ correlations are expected to be
sensitive to the amount of spin-orbit splitting. To quantify this
statement we have repeated our SMMC studies of $^{50}$Fe and $^{56}$Fe
by artificially reducing the spin-orbit splitting in the
single-particle energies by a factor of 2; thus we use
$\epsilon_{7/2}=0.,\epsilon_{3/2}=2.,
\epsilon_{5/2}=3.,\epsilon_{1/2}=3.$ As expected, the reduction of the
spin-orbit splitting generally increases the isoscalar correlations at
the expense of the isovector correlations. In particular, the
isoscalar correlations due to the coupling of the spin-orbit partners
$f_{7/2}-f_{5/2}$ are strongly increased (by about a factor of 2.5 in
these nuclei). However, a deviation from this general trend is also
noteworthy. The increase of isoscalar correlations leads also to an
increase of $nn$ correlations in $^{56}$Fe due to increase of
correlations in the $f_{5/2}$ orbital.

Obviously coupling between $f_{7/2}f_{5/2}$ is very strong. Will it
dominate if the spin-orbit splitting were removed? Unfortunately such
a calculation cannot be performed with the SMMC method due to the
breakdown of the g-extrapolation~\cite{sign}. For that reason we have
gone back to $^{46}$V and have performed one shell model
diagonalization study with the KB3 interaction and one in which the
monopole terms have been removed from the interaction.  In both cases
we then calculated the expectation value of the isovector and
isoscalar pairing Hamiltonians in the lowest $T=0$ state, which is the
ground state in this academic case.  For the
eigenstate of the realistic Hamiltonian we find 
$\langle H_{pair}^{T=1} \rangle =2.18$~MeV and 
$\langle H_{pair}^{T=0} \rangle =1.41$~MeV,
while the magnitude is inverted if the monopole terms are taken out:
$\langle H_{pair}^{T=1} \rangle =2.09$ MeV 
and $\langle H_{pair}^{T=0} \rangle =4.62$ MeV. It is thus obvious that
isovector correlations in odd-odd $N=Z$ nuclei win over isoscalar
correlations due to the presence of the spin-orbit splitting.

\section{conclusions}

Modern diagonalization codes make it possible to study nuclear spectra
state-by-state even for nuclei rather far away from closed shells,
where correlations play a decisive role. Here, we use the $N = Z$
odd-odd nucleus $^{46}$V as a case study for the investigation of
pairing and other correlations.  We are able to describe not only the
ground state, but the development of various correlations with the
excitation energy and, respectively, temperature.

Our main emphasis is on the competition between the two basic modes
(like-particle ($pp$ and $nn$) and proton-neutron ($pn$)) of the
isovector pairing on one hand, and the isoscalar pairing on the other
hand. These features are particularly important in $^{46}$V, where the
ground state spin and isospin $J,T = 0,1$ demonstrates the importance
of the $np$ isovector pairing.  We show that this mode of pairing
peaks sharply at low energies.  At energies or temperatures above the
isovector pairing gap ($E \ge 3$ MeV or $T \ge 0.5$ MeV) the three
isovector pairing strengths become essentially equal to each other,
and gradually decrease. At the same time, the isoscalar pairing
energy, which is smaller than the isovector one inside the gap,
becomes at higher energies comparable with the $total$ isovector
pairing, or even bigger.
 
We have shown (Fig. 2) that the large statistical weight of the
low-lying excited isospin $T = 0$ states causes a characteristic
maximum in the specific heat $C_V$. This feature represents an analog
of the phase transition from the $T=1$ ground state to the regime
where the partition function and internal energy are dominated by the
$T=0$ states.

Finally, we study the contributions to the correlation energy at different
temperatures in $^{46}$V. We show, first of all, that as expected
pairing and quadrupole correlations account for most of the total
correlation energy everywhere. All other correlations contribute
about 2 MeV, and this amount is essentially independent of temperature.
Pairing dominates near the ground state, but soon 
(for temperatures above 250 keV) pairing and quadrupole
correlation energies become comparable.

In an additional study based on the SMMC we show how pair correlations
in Fe isotopes depend on $| N - Z|$. In particular, we were able to
trace the contribution of individual subshells to the pairing
strength. That calculation also confirmed that the isovector $pn$
pairing strength decreases fast with increasing $| N - Z|$. On the
other hand, the isoscalar $pn$ pairing in this chain of isotopes is
almost independent on $| N - Z|$.

In the SU(4) limit, i.e. for degenerate single particle levels,
identical coupling constant of the isoscalar and isovector pairing
force and absence of other components in the residual interaction, the
lowest $T=0$ and $T=1$ states are degenerate. However, these
symmetries are lifted in realistic cases and, for the present nucleus
$^{46}$V, the ground state has $T=1$. At first we note that the
coupling constant of the isoscalar and isovector pairing force,
determined in Ref.~\cite{zuker}, are $G_{p01}$ = 0.37 MeV, and
$G_{p10}$ = 0.57 MeV, i.e. the isoscalar pairing is stronger.  Why
does the isovector pairing nevertheless dominate the ground state? We
argued that the effect of the isoscalar pairing is weakened by the
spin-orbit splitting which introduces additional energy scale into the
problem.  The isoscalar pairing can act efficiently only at energies
where the spin-orbit splitting plays only a minor role. To prove that
point we have artificially reduced the spin orbit splitting in Fe
isotopes, and indeed observed an increase of the isoscalar pairing.
But through higher order effects the isovector pairing was affected as
well. Furthermore, we have shown that in the lowest $T=0$ state in
$^{46}$V the isoscalar correlations would prevail if the spin-orbit
force were switched off. Thus, this somewhat academic exercise has
confirmed to us that without the strong spin-orbit force the nuclear
ground states would look very differently from the ones we are
familiar with.

\acknowledgements

This work was supported in part by the Danish Research Council and by
the U.S. Department of Energy under grant DE-FG03-88ER-40397.

\end{multicols}

\begin{multicols}{2}
\narrowtext

\begin{figure}
  \begin{center}
    \leavevmode
    \epsfxsize=0.8\columnwidth
    \epsffile{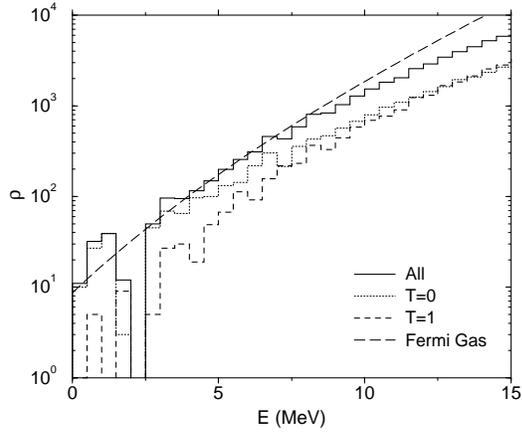}
    \caption{Level density of $^{46}$V as function of the excitation
      energy. The levels with isospin $T$=0 and 1 are shown
      separately. The Fermi gas level density parameters were $a =
      3.5$~MeV$^{-1}$ and $\Delta = -2.0$~MeV.}
    \label{fig3}
  \end{center}
\end{figure}

\begin{figure}
  \begin{center}
    \leavevmode
    \epsfxsize=0.8\columnwidth
    \epsffile{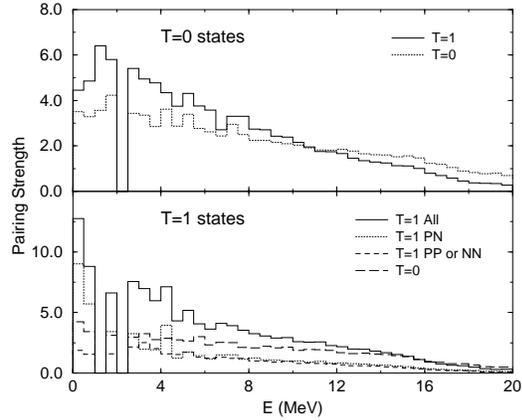}
    \caption{Average isovector $T = 1, S = 0, L = 0$ and isoscalar $T = 0, S = 
      1, L = 0$ pair correlation strengths in $^{46}$V as function of
      the excitation energy (binned for clarity in bins 0.5 MeV
      wide).}
    \label{fig4}
  \end{center}
\end{figure}

\begin{figure}
  \begin{center}
    \leavevmode
    \epsfxsize=0.8\columnwidth
    \epsffile{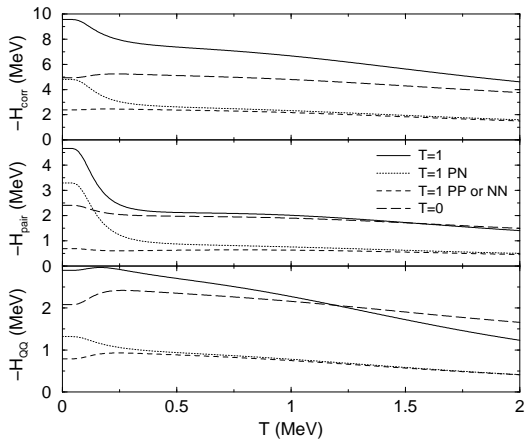}
    \caption{Expectation values of the total correlation
      energy, isoscalar and isovector parts of pairing, and
      quadrupole-quadrupole energies as a function of the temperature
      for $^{46}$V.}
    \label{fig5}
  \end{center}
\end{figure}

\begin{figure}
  \begin{center}
    \leavevmode 
    \epsfxsize=0.8\columnwidth 
    \epsffile{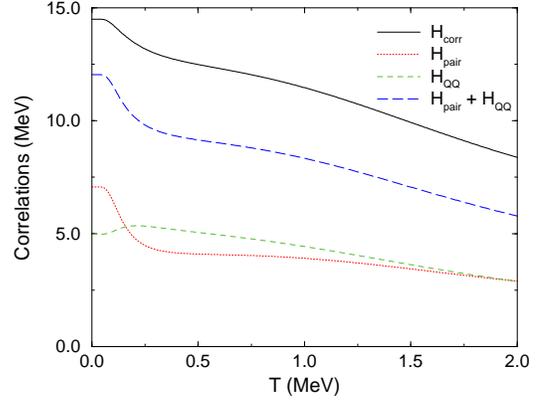}
    \caption{Expectation values of the pairing, quadrupole-quadrupole and
      total correlation energies as a function of the temperature for
      $^{46}$V.}
    \label{fig6}
  \end{center}
\end{figure}

\begin{figure}
  \begin{center}
    \leavevmode 
    \epsfxsize=0.9\columnwidth 
    \epsffile{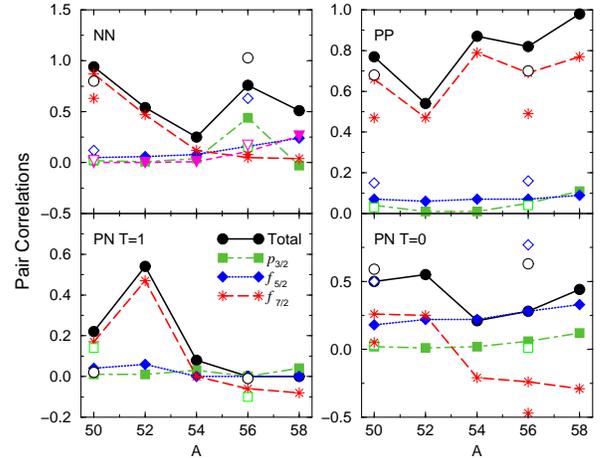}
    \caption{Isovector and isoscalar pairing correlation in Fe
      isotopes. The excess over the mean field value is plotted.  The
      unfilled symbols refer to the SMMC studies of $^{50}$Fe and
      $^{56}$Fe, in which we have artificially halfed the spin-orbit
      splitting in the single particle energies (see text).  }
    \label{fig7}
  \end{center}
\end{figure}

\end{multicols}


\begin{references}
  
\bibitem{Engel96} J. Engel, K. Langanke and P.Vogel, Phys. Lett. B
  {\bf 389}, 211 (1996).
  
\bibitem{Langanke97} K. Langanke, D.J. Dean, P.B. Radha and S.E.
  Koonin, Nucl. Phys. A {\bf 613}, 253 (1997).
  
\bibitem{Satula97} W. Satula and and R. Wyss, Phys. Lett. B {\bf 393},
  1 (1997).
  
\bibitem{Civitarese97} O. Civitarese and M. Reboiro, Phys. Rev. C {\bf
    56}, 1781 (1997); O. Civitarese, M. Reboiro, and P. Vogel, Phys.
  Rev. C {\bf 56}, 1840 (1997).
    
\bibitem{Dobes97} J. Dobe\v{s}, Phys. Lett. B {\bf 413}, 239 (1997).
  
\bibitem{Poves98} A. Poves and G. Mart\'{\i}nez-Pinedo, Phys. Lett. B
  {\bf 430}, 203 (1998).
  
\bibitem{Engel97} J. Engel, S. Pittel, M. Stoitsov, P.Vogel and J.
  Dukelsky, Phys. Rev. C {\bf 55}, 1781 (1997).
  
\bibitem{Engel98} J. Engel, K. Langanke and P. Vogel, Phys. Lett. B
  {\bf 429}, 215 (1998).
  
\bibitem{VanIsacker97} P. Van Isacker and D.D. Werner, Phys. Rev. Lett.
  {\bf 78}, 3266 (1997).
  
\bibitem{Vogel98} P. Vogel, nucl-th/9805015; J. J\"{a}necke, Nucl. Phys.
{\bf 73}, 97 (1965).
  
\bibitem{Satula98} W. Satula {\em et al.}, Phys. Lett. B {\bf 407}
  (1997) 103

\bibitem{Rudolph} W. Rudolph {\em et al.}, Phys. Rev. Lett. {\bf 76},
  376 (1996).
  
\bibitem{Dean97} D.J. Dean, S.E. Koonin, K. Langanke and P.B. Radha,
  Phys. Lett. B {\bf 399}, 1 (1997).
  
\bibitem{Poves81} A. Poves and A.\,P. Zuker, Phys. Rep. {\bf 70}, 235
  (1981). 
  

\bibitem{Antoine} E. Caurier, computer code ANTOINE, CRN, Strasbourg,
  1989. 
  
\bibitem{Caurier98} E. Caurier, G. Mart\'{\i}nez-Pinedo, F. Nowacki,
  A. Poves, J. Retamosa, A.\,P. Zuker, LANL archive nucl-th/9809068. 
  
\bibitem{Johnson} C.W. Johnson, S.E. Koonin, G.H. Lang and W.E.
  Ormand, Phys. Rev. Lett. {\bf 69}, 3157 (1992).

\bibitem{Lang} G.H. Lang, C.W. Johnson, S.E. Koonin and W.E. Ormand,
  Phys. Rev. {\bf C48}, 1518 (1993).
  
\bibitem{physrep} S.E. Koonin, D.J. Dean and K. Langanke, Phys.
  Rep. {\bf 278}, 1 (1997).

\bibitem{sign} Y. Alhassid, D.J. Dean, S.E. Koonin, G. Lang and W.E.
  Ormand, Phys. Rev. Lett. {\bf 72}, 613 (1994).
  
\bibitem{Caurier94} E. Caurier, A.\,P. Zuker, A. Poves, and
  G.~Mart\'{\i}nez-Pinedo, Phys. Rev. C {\bf 50}, 225 (1994).
  
\bibitem{nds} L. K. Peker, Nucl. Data Sheets {\bf 68}, 1 (1993).
  
\bibitem{Thielemann} J.J. Cowan, F.-K. Thielemann and J.W. Truran,
  Phys. Rep. {\bf 208}, 267 (1991).
  
\bibitem{Woosley} S. E. Woosley, W. A. Fowler, J. A. Holmes, and B. A.
  Zimmerman, At. Nucl. Data Tables {\bf 22}, 371 (1978).

\bibitem{Bes} D.R. Bes and R.A. Sorensen, Adv. Nucl. Phys. {\bf 2},
  129 (1969).
  
\bibitem{french} J.\,B. French, in {\em Isospin in Nuclear Physics},
  edited by D.\,H. Wilkinson (North-Holland, Amsterdam, 1969).
  
\bibitem{zuker} M. Dufour and A.\,P. Zuker, Phys. Rev. C {\bf 54},
  1641 (1996).
  
\end{references}
\end{document}